\documentclass[12pt,onecolumn]{IEEEtran}
\usepackage{algorithm,algorithmic,amsbsy,amsmath,amssymb,epsfig,bbm,mathrsfs,multirow,amsthm}
\usepackage{graphicx,caption,subcaption,array,multirow}
\usepackage{hyperref}
\hypersetup{
	colorlinks=true,
	linkcolor=blue,
	filecolor=blue,   
	urlcolor=black,citecolor=blue,}

\begin{document}

\title{``Borrowing Arrows with Thatched Boats'': The Art of Defeating Reactive Jammers in IoT Networks}

\author{Dinh Thai Hoang, Diep N. Nguyen, Mohammad Abu Alsheikh, Shimin Gong, \\ Eryk Dutkiewicz, Dusit Niyato, and Zhu Han}


\maketitle

\begin{abstract}
In this article, we introduce a novel deception strategy which is inspired by the ``Borrowing Arrows with Thatched Boats'', one of the most famous military tactics in the history, in order to defeat reactive jamming attacks for low-power IoT networks. Our proposed strategy allows resource-constrained IoT devices to be able to defeat powerful reactive jammers by leveraging their own jamming signals. More specifically, by stimulating the jammer to attack the channel through transmitting fake transmissions, the IoT system can not only undermine the jammer's power, but also harvest energy or utilize jamming signals as a communication means to transmit data through using RF energy harvesting and ambient backscatter techniques, respectively. Furthermore, we develop a low-cost deep reinforcement learning framework that enables the hardware-constrained IoT device to quickly obtain an optimal defense policy without requiring any information about the jammer in advance. Simulation results reveal that our proposed framework can not only be very effective in defeating reactive jamming attacks, but also leverage jammer's power to enhance system performance for the IoT network. 
\end{abstract}


\section{Introduction}
\label{sec:introduction}

Over the last 10 years, we have witnessed an explosive growth of Internet-of-Things (IoT) applications with great influences in many sectors such as manufacturing, healthcare, smart cities, and industry 4.0~\cite{Fuqaha2015Internet}. The development of IoT has brought great benefits to human life and opened many potential market opportunities for equipment manufacturers, Internet service providers and application developers. According to IDC's prediction, worldwide technology spending on the IoT can reach \$1.2T in 2022, attaining a compound annual growth rate of 13.6\% over the 2017-2022 forecast period\footnote{\href{https://www.idc.com/getdoc.jsp?containerId=prUS43994118}{\text{IDC Forecasts Worldwide Technology Spending on the Internet of Things to Reach \$1.2 Trillion in 2022}}}. Furthermore, it is also estimated that around 500 billion devices will be connected to the Internet by 2030\footnote{\href{https://www.cisco.com/c/dam/en/us/products/collateral/se/internet-of-things/at-a-glance-c45-731471.pdf}{\text{Internet of Things At a Glance - Cisco}}}. Obviously, with outstanding advantages and benefits, IoT has been becoming an indispensable part of human life in the near future.



Despite the explosive growth, IoT is extremely vulnerable to security threats, especially jamming attacks, due to hardware constraints and the broadcast nature of wireless communications. In particular, by transmitting high-power jamming signals to a target channel, a jammer can degrade Signal-to-Interference-plus-Noise Ratio (SINR) at the IoT receiver, e.g., an IoT gateway. Consequently, the IoT receiver is unable to decode information from the IoT transmitter. More importantly, radio jamming attacks can be easily launched by using commercial off-the-shelf products\footnote{\url{www.jammer-store.com}}, and thus they can cause serious consequences to human life, especially in mission-critical sectors such as healthcare, military, and transportation. For example, an attacker used a cheap jamming device to perform a car lock jamming attack, with the intent of breaking into vehicles, caused chaos in a parking lot where nobody could unlock/lock their remote car locks and ended up triggering the number of alarms in the process~\cite{Manchester}. As a result, solutions to deal with jamming attacks are of urgent needs for future development of IoT networks.


In this paper, we first give an overview about communication methods and potential vulnerabilities to jamming attacks in IoT networks. We then review emerging wireless jamming techniques and current effective countermeasures to defeat jamming attacks. After that, we develop a novel anti-jamming strategy which allows resource-constrained IoT devices to effectively to defeat powerful reactive jammers. Our proposed strategy is inspired by a famous deception military tactic, called \emph{Borrowing Arrows with Thatched Boats}\footnote{\href{https://www.theepochtimes.com/chinese-idioms-borrowing-arrows-with-thatched-boats-\%E8\%8D\%89\%E8\%88\%B9\%E5\%80\%9F\%E7\%AE\%AD_309040.html}{\text{Chinese Idioms: Borrowing Arrows With Thatched Boats}}}, which utilizes the enemy's power to defeat the enemy itself. In particular, considering an IoT system under jamming attacks by a reactive jammer, the IoT transmitter can perform deception strategies by sending fake transmissions to its receiver to stimulate the jammer to attack the channel. Once the jammer attacks the channel, the IoT transmitter can immediately harvest energy from jamming signals or leverage the jamming signals as a communication means to transmit data to its receiver by using ambient backscatter techniques~\cite{Huynh2018Ambient}. In this way, we can not only undermine the jammer's power, but also utilize the jammer as an additional energy source to improve performance for the IoT system. In addition, a low-cost deep reinforcement learning (DRL) algorithm is developed to quickly obtain an optimal deception policy of the anti-jamming strategy for the IoT system without requiring any information about the jammer in advance. The simulation results then clearly show the efficiency of our proposed framework in terms of higher throughput and lower dropped packets in dealing with reactive jamming attacks. 


\section{Communications in IoT Networks and Vulnerabilities to Jamming Attacks}
\label{sec:Overview}

\subsection{IoT Communication Protocols: A Brief Overview}

In order to connect billions of smart devices to the Internet, wireless communications have been widely adopted as the most effective communication medium in IoT networks. In practice, typical communication protocols used in IoT networks are WiFi, Bluetooth, IEEE 802.15.4, Z-wave, and LTE-Advanced~\cite{Fuqaha2015Internet}. Furthermore, there are some other wireless communications technologies introduced recently for IoT networks such as Radio-Frequency IDentification (RFID), Near Field Communication (NFC), Narrowband IoT (NB-IoT) and Long Range (LoRa). Each protocol possesses its own advantages as well as limitations, and usually it is designed to serve for a particular type of IoT applications. For example, IEEE 802.15.4 protocol is usually used for indoor ultra-low power IoT communication applications, while LTE-Advanced protocol is typically implemented for outdoor low-latency IoT communication applications.

\subsection{Vulnerabilities of IoT Communications to Jamming Attacks}

In practice, IoT networks are allocated to some particular channels for communications. For example, according to the newest IEEE communication standard for low-rate networks released in 2017, there are still only $16$ channels allocated for all IoT devices using the IEEE 802.15.4 communication protocol in $2.4$GHz band ($2400$-$2483$MHz)\footnote{IEEE Standard 802.11.4t$^\textnormal{TM}$-2017.}. In addition, due to hardware constraints and power supply limitations, transmit powers of IoT devices are usually kept at very low levels. For example, IoT devices using IEEE 802.15.4 communication protocol usually have transmit powers at approximately $0$ dBm~\cite{Yuan2013Coexistence}. These characteristics, i.e., the limited number of communication channels and low transmit power, are especially susceptible to jamming attacks. In particular, an attacker can easily determine a target IoT communication channel, and then transmit high-power jamming signals to prevent communications on the channel. As a result, the IoT receiver cannot successfully decode information from the IoT transmitter due to very low SINR. More importantly, conventional security methods, e.g., encryption, cannot be used to defeat jamming attacks.

\subsection{Radio Jamming Attacks in IoT Networks}

A radio jamming attack is a method in which a jammer intentionally transmits strong radio jamming signals to the target channel to illegally interrupt IoT communications by decreasing the SINR at the IoT receiver. In the past, radio jamming attack is one of the most common technology methods used during World War II to interfere radar operations used to guide enemies' missiles/aircrafts or prevent citizens from listening foreign radio stations broadcast by enemies. In the modern life, radio jamming attacks are more and more dangerous as they can be maliciously used not only in military but also civil applications. In particular, with the explosion of IoT applications in many areas in our life, e.g., smart home, manufacturing, healthcare, and transportation, jamming attacks can cause damages to IoT systems, leading to serious adverse impacts to the human life. For example, an attacker could easily disable the smart home wireless alarm system by using a commercial portal jamming device to prevent the IoT devices from sending alarm messages to the server and/or the home owner. More importantly, in some sensitive IoT applications, e.g., brain and heart implanted IoT devices, jamming attacks can cause serious consequences for human health. Therefore, it is an urgent call for countermeasures to defeat jamming attacks in IoT networks. 

There are some typical methods used in radio jamming attacks~\cite{Mpitziopoulos2009Asurvey}. 
\begin{itemize}
	\item \emph{Constant jamming}: A jammer continuously emits meaningless noise signals on the target channel without following any standard protocol.
	\item \emph{Deceptive jamming}: A jammer continuously transmits regular packets to the channel instead of emitting random bits as the constant jamming attack, to make the victim to believe that a legitimate device is transmitting. 
	\item \emph{Random jamming}: A jammer alternates between sleeping and jamming modes to save energy. In the jamming mode, the jammer can perform either constant or deceptive jamming attacks.
	\item \emph{Reactive jamming}: A jammer listens the target channel and transmits jamming signals (either constant or deceptive signals) to the channel once it detects activities of the victim on the channel.
\end{itemize}	

Among radio jamming methods, reactive jamming is the most effective and serious attack in IoT networks as it can ``smartly'' detect and launch attacks only when the IoT system is active. Furthermore, it is more difficult to detect a reactive jamming than a proactive jamming, e.g., constant and deceptive jamming, because it is only active when the IoT transmitter is transmitting, and thus a detector might not be able to distinguish whether signals on the channel are from the IoT transmitter or from a jammer. More seriously, reactive jamming devices can be developed easily by integrating compact signal-detection circuits on conventional jamming devices. Many commercial reactive jamming devices have been developing for military and civil purposes, e.g., Roshel (\url{www.roshel.ca}) and TJ Infotech (\url{www.tjinfotech.net}), and this makes defeating jamming attacks in IoT networks more challenging than ever.


\section{Solutions to Defeat Radio Jamming Attacks in IoT Networks}
\label{sec:Solutions}

\subsection{Frequency-Hopping}

This is one of the most common methods used to deal with radio jamming attacks. The key idea of this method is that if the current communication channel is attacked, the IoT devices will find another channel that is not being attacked for communication. This solution is effective to defeat proactive jamming attacks as the IoT devices have information about channels under being attacks in advance. However, this solution requires a set of available communication channels, which are an expensive and scarce resource, together with the a predefined, but preferably adaptive channel-switching algorithm implemented on all the IoT devices in advance. As a result, this solution might not be effective to widely implement on hardware-constrained and channel-limited IoT communication systems. In addition, this solution cannot deal with reactive jamming attacks in which the jammers only launch attacks after the IoT devices transmit signals.

\subsection{Rate Adaptation}

Rate adaptation is also an anti-jamming method which can be effectively implemented in IoT networks to defeat proactive jamming attacks. For this method, when a channel is under a proactive jamming attack, the IoT transmitter will observe the channel status and determine an appropriate transmission rate such that the IoT receiver still can decode information under the jamming attack. This method can be implemented on a hardware-constrained IoT device as the IoT device only needs to control its data transmission rate, typically at a low rate, when the attack occurs. However, this method does not work well if the jammer always attacks the channel at high power levels. In addition, similar to the frequency-hopping approach, this method cannot deal with reactive jammers since they only attack the channel after detecting signals of the IoT transmitter on the channel.

\subsection{Ambient Backscatter}

Thanks to the development of backscatter communication techniques, there is a new anti-jamming approach introduced recently~\cite{HuynhJam2019}. This approach is developed based on the idea of recent advanced ambient backscatter communication technologies~\cite{Liu2013Ambient}~\cite{Kimionis2014Increased}. Specifically, in an ambient backscatter communication system, a transmitter can transmit data to its receiver by backscattering surrounding RF signals, e.g., TV or FM signals, to its receiver. The receiver then can decode the information by using some low-cost decoding techniques, e.g., averaging mechanisms~\cite{Liu2013Ambient}. Inspired by this idea, the authors in~\cite{HuynhJam2019} proposed a solution utilizing jamming signals as a ``means'' to delivery data for IoT communication systems. In particular, when a jammer transmits jamming signals to the channel, the IoT transmitter can leverage the jamming signals to backscatter and transmit information to the IoT receiver. In this way, the IoT system can not only avoid the jamming attacks, but also utilize the jamming signals for its transmissions. As a result, this approach is very effective in dealing with jamming attacks in IoT networks. However, similar to all aforementioned methods, this approach still fails to defeat the reactive jamming attacks. The reason is that this approach is based on jamming signals; however, for reactive jammers, they only attack the channel if they detect signals transmitted by the IoT device on the channel.

In practice, reactive jamming is the most difficult attack to deal with especially in low-power IoT networks. Some solutions were proposed to combat reactive jamming attacks such as~\cite{Sciancalepore2018Strength,Xu2012Joint}, but they all rely on the assumptions that there are multiple communication channels and jammers cannot attack all the channels simultaneously. To the best of our knowledge, there is still no effective solution to deal with reactive jamming attacks, and this is the motivation for us to introduce Deep$Q$Fake, a novel framework to defeat jamming attacks in IoT networks.

%
%

\section{Deep$Q$Fake: A DRL-based Deception Strategy to Defeat Reactive Jamming Attacks}
\label{sec:framework}

\subsection{Deception Strategy to Defeat Reactive Jamming Attacks}
\label{subsec:}

\subsubsection{Functions and components of the IoT system}

\begin{figure*}[!]
	\centering
	\includegraphics[width=\linewidth]{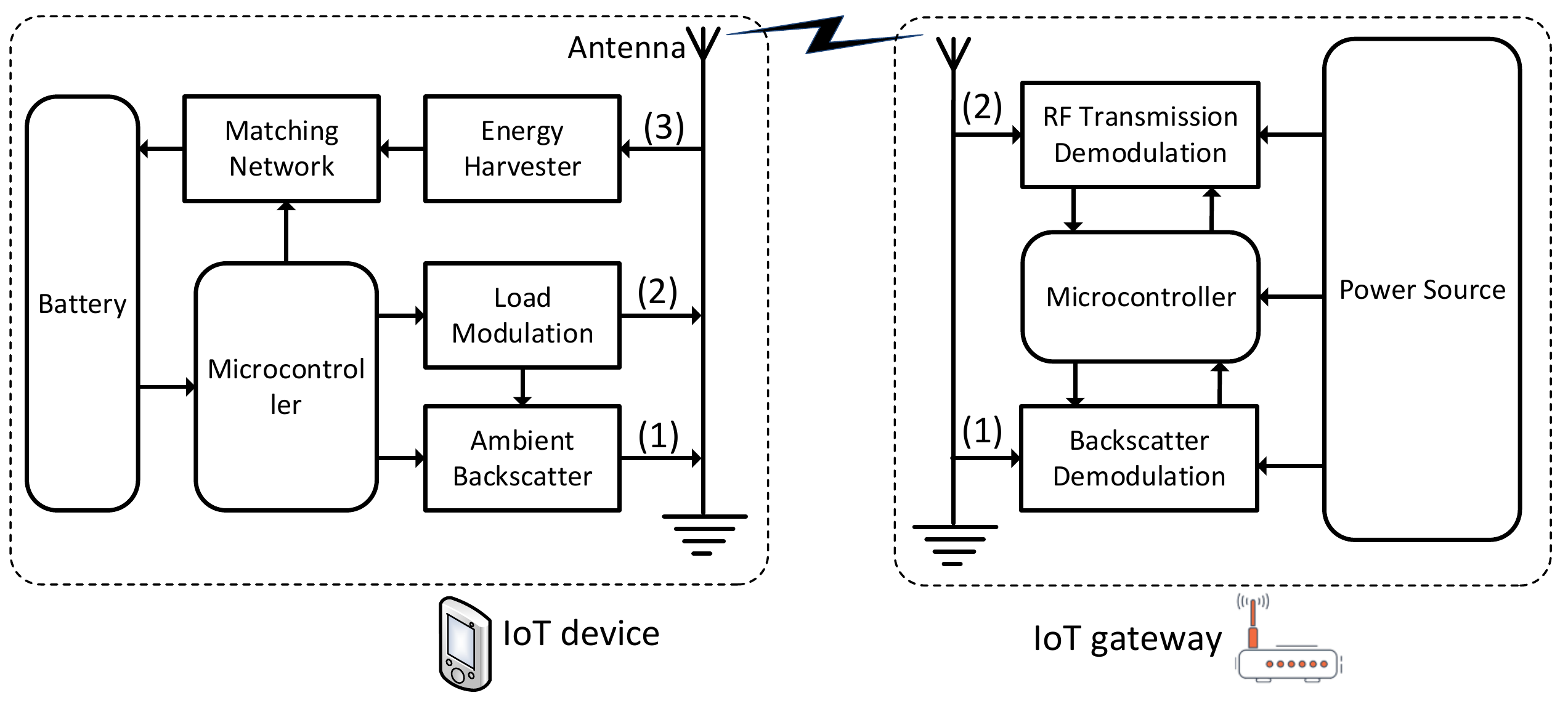}
	\caption{Circuit diagram of the low-power IoT system.}
	\label{fig_IoT_Architecture}
\end{figure*}

In this paper, we consider an IoT communication system including a low-power IoT device and an IoT gateway. The IoT device is equipped with a storage to store data and a battery to store energy. The energy stored in the battery will be used to transmit data in the storage to the IoT gateway. As a low-power device, the IoT device is equipped with an RF energy harvesting circuit and an ambient backscatter circuit. The energy harvesting circuit is used to harvest energy from surrounding environment and store the harvested energy in the battery. Meanwhile, the ambient backscatter circuit is used in the case when the IoT device wants to transmit data to the gateway based on the ambient backscatter technique. Both circuits share the same antenna and are under control by a microcontroller as illustrated in Fig.~\ref{fig_IoT_Architecture}. In practice, energy harvesting and ambient backscatter circuits are tiny and consume a small amount of energy which are especially suitable to implement on the low-power IoT device. For example, RFD102A (RF-DC Converter) module has the size of only 5mm$\times$7mm$\times$1.8mm, and its energy harvesting efficiency can reach up to 50\%. Furthermore, ADG902 RF switch used in ambient backscatter circuit is even smaller than RFD102A, and its energy consumption for backscatter mode is very small with only 0.25$\mu$W. More details about components and functions of energy harvesting and ambient backscatter circuits can be found in~\cite{Liu2013Ambient,Huynh2018Ambient}.

There are three main functions on the IoT device as shown in Fig.~\ref{fig_IoT_Architecture}. First, the IoT device can use the ambient backscatter circuit to transmit data to the gateway by backscattering surrounding RF signals. In this case, the gateway will use corresponding backscatter demodulator circuit, e.g., a low-cost averaging circuit~\cite{Liu2013Ambient}, to decode and extract information. Second, the IoT device can actively transmit data to the gateway by using energy in the battery. In this case, the gateway can use the conventional demodulator to decode information. Finally, the third function is to harvest energy from surrounding environment. It is important to note that only one function can be performed at a time due to single antenna usage. Thus, energy and communication need to be optimized to maximize the network throughput for IoT system~\cite{Huynh2018Ambient}.

\subsubsection{Reactive jammer and intelligent deception strategy}

\begin{figure*}[!]
	\centering
	\includegraphics[width=\linewidth]{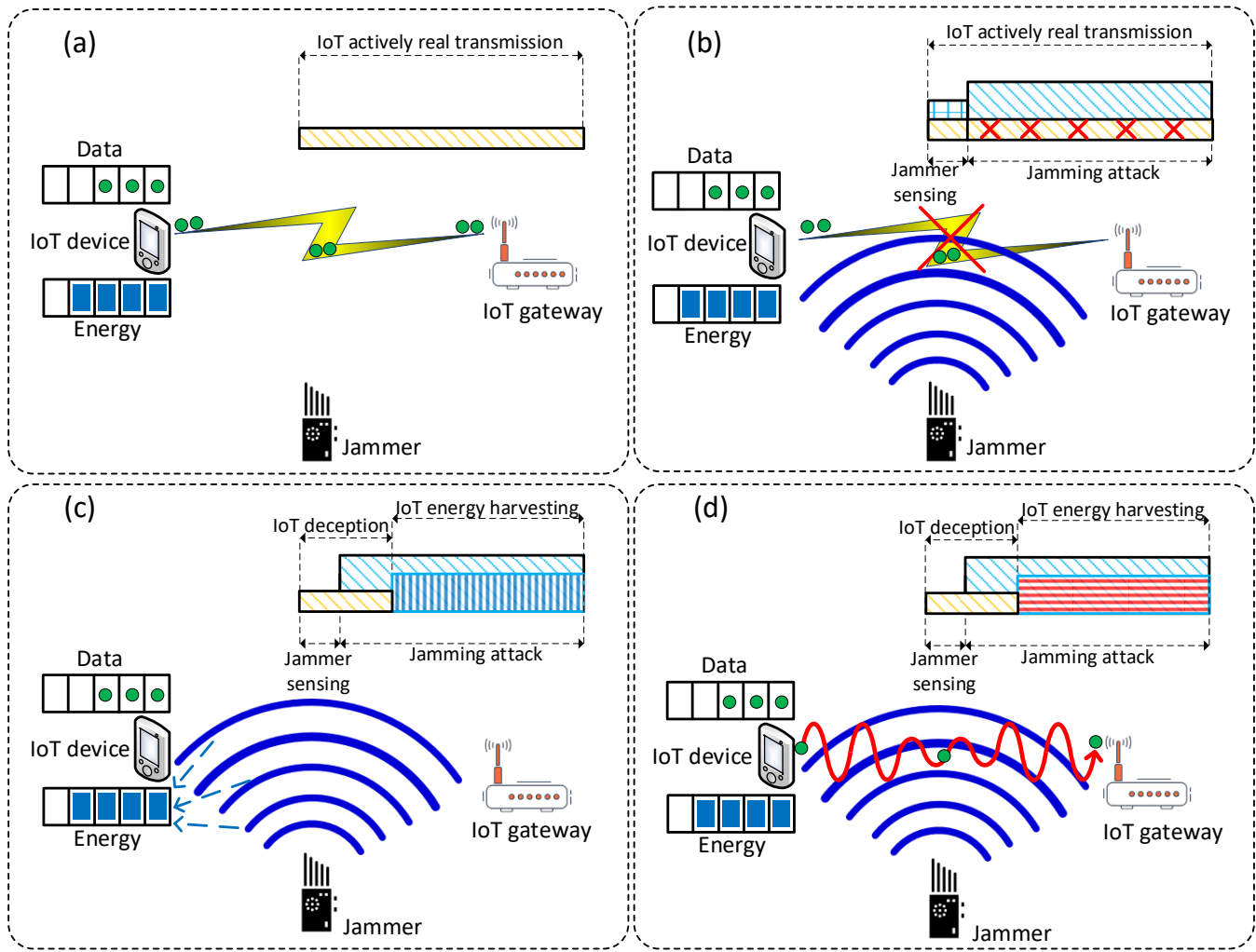}
	\caption{System model.}
	\label{fig_System_Model}
\end{figure*}

In this work, we consider a smart and reactive jammer which only attacks the channel if it detects signals transmitted from the IoT transmitter. The jammer is equipped with a detector circuit, e.g., matched filter or energy detector, to detect activities of the IoT device on the channel. This circuit will listen to the channel and identify whether the signal on the channel is from the IoT device or not. If the signal is determined from the IoT transmitter and if the jammer has sufficient energy to launch an attack, the jammer will immediately transmit jamming signals at high power to the channel as illustrated in Fig.~\ref{fig_System_Model}(b). The aim of jammer is to significantly increase the noise at the gateway, thereby preventing the gateway from successfully decoding information from the IoT device. In the case if the IoT device actively transmits data and the jammer does not have sufficient energy to launch an attack, then the IoT gateway can successfully decode information from the IoT transmitter as illustrated in Fig.~\ref{fig_System_Model}(a).

In practice, the jammer's energy is much stronger than that of IoT device, and thus the jammer can attack and prevent all communications from the IoT device. To deal with reactive and strong jamming attacks, we introduce an intelligent deception strategy for the IoT system. In particular, instead of always performing actual transmissions, the IoT device can ``sometimes'' send ``fake'' transmissions to lure the jammer. A fake transmission is defined to be a deception strategy in which the IoT device only transmits signals for a short period of time to attract the jammer. By using the deception mechanism, we can undermine the attack ability of jammer, and thus the jammer may not be able to attack when the IoT device performs actual transmissions. More importantly, by luring the jammer to transmit jamming signals, the IoT device can utilize these signals to enhance its energy and communication efficiency. Specifically, when the IoT device performs a fake transmission and the jammer attacks the channel, we introduce two intelligent mechanisms to ``exploit'' jamming signals as follows:
\begin{itemize}
	\item \emph{Deception then harvest energy}: For this scheme, the IoT device transmits signals on the channel for a short period of time at the beginning of a time slot. Then, if the jammer performs jamming attacks, the IoT device will harvest energy from jamming signals until the jammer stops jamming. This scheme is illustrated in Fig.~\ref{fig_System_Model}(c).
	\item \emph{Deception then backscatter data}: Similar to the previous deception scheme, the IoT device also transmits signals for a short period of time at the beginning of a time slot to lure the jammer. Then, if the jammer performs jamming attacks, the IoT device will leverage the jamming signals to backscatter data to the gateway. This scheme is illustrated in Fig.~\ref{fig_System_Model}(d).
\end{itemize}

It can be clearly seen that the proposed deception strategies are very effective to deal with reactive jamming attacks. First, the energy harvesting function is especially useful for the low-power IoT device in harvesting energy not only from surrounding environment but also from jamming signals. The harvested energy will be used to transmit data to the gateway and support operations at the IoT device. Similarly, ambient backscatter function is also very useful in supporting free-cost data transmission for the IoT device by reflecting RF signals from surrounding environment or jamming signals. In~\cite{Liu2013Ambient}, the authors show that ambient backscatter can be used for the communications between two bateryless IoT devices. More interestingly, these two functions are even more effective under strong jamming attacks. Intuitively, the more power the jammer uses to attack the channel, the larger amount of energy the IoT device can harvest and the more bits the IoT system can successfully backscatter. These results have been verified based on information theoretic approaches as well as many experiments as shown in the recent survey~\cite{Huynh2018Ambient}. 

Although the aforementioned deception strategies clearly benefit the IoT system in dealing with reactive jamming attacks, they can only perform best when they have some information about the jammer in advance. For example, given the jamming signal on the channel, the IoT device should harvest energy or perform an ambient backscatter transmission from the jamming signal? In addition, how to optimize energy harvesting, active transmission, and backscatter transmission processes without knowing jammer's capacity, e.g., frequency and power of attacks, in advance? The jammer is a malicious device which is used to prevent IoT communications, and thus its information is nearly impossible to obtain in advance. Therefore, to deal with the challenges in finding the optimal deception strategy for the IoT device without requiring the knowledge about the jammer in advance, in the next section, we introduce Deep$Q$Fake, a DRL-based deception framework. This framework allows the IoT device to learn the jammer's strategy through real-time interactions. 

\subsection{Deep$Q$Fake: A DRL-based Deception Strategy}

\subsubsection{Reinforcement Learning}

Reinforcement learning is one of important branches of machine learning which has been widely implemented in practice for real-time decision making problems when agents, e.g., IoT devices, do not have sufficient information about its surrounding environment. In a reinforcement learning process, an agent interacts with its surrounding environment through trial-and-error processes, i.e., making decisions and observing results, then gradually it can learn the characteristics of environment and make ``intelligent'' decisions to maximize its long-term reward as illustrated in Fig.~\ref{fig_RLvsDRL}(a). As a result, reinforcement learning has been successfully applied in many critical sectors such as robotics, economic, artificial intelligent (AI), and manufacturing~\cite{Sutton1998Reinforcement}. Thus, this is motivation for us to introduce an application of reinforcement learning to help the IoT device (i.e., the agent) to learn characteristics of jammer (i.e., surrounding environment) in order to maximize the average throughput of IoT system (i.e., long-term reward) without requiring the jammer's information (e.g., frequency and power of attacks) in advance. 

In order to maximize efficiency of the learning process, the first important step is to determine appropriate states, actions, and rewards for the IoT system. In the considered system, we aim to maximize the long-term throughput for IoT system, and thus the reward of learning process can be defined to be the average IoT system throughput. In addition, to achieve this objective, the IoT device needs to determine the best action to perform at each time slot. There are four possible actions, i.e., IoT operation modes, for the IoT device to choose from at each time slot.
\begin{itemize}
	\item \emph{Passive Energy Harvesting:} In this mode, the IoT device listens the channel and harvests energy from surrounding RF signals if available. This mode is used when the IoT device has no data to transmit or no energy for active transmissions. 
	\item \emph{Active Transmission:} In this mode, the IoT device proactively uses its energy to transmit data to the gateway. This mode is the basic communication function of almost all current IoT devices. The advantage of this mode is easy to implement and efficient in transmitting the number of packets to the gateway as illustrated in Fig.~\ref{fig_System_Model}(a). However, if a jamming attack is launched when the IoT device uses this mode, all packets will be corrupted as illustrated in Fig.~\ref{fig_System_Model}(b). 
	\item \emph{Active Deception and Energy Harvesting:} This mode allows the IoT device to perform a fake transmission to lure the jammer. Then, if the jammer attacks the channel, the IoT device will actively harvest energy from the jamming signals as illustrated in Fig.~\ref{fig_System_Model}(c). 
	\item \emph{Active Deception and Information Backscattering:} This mode allows the IoT to perform a deception transmission to lure the jammer. Then, if the jammer attacks the channel, the IoT device will actively backscatter the jamming signals to transmit data to the gateway as illustrated in Fig.~\ref{fig_System_Model}(d). 
\end{itemize}

Note that, in the case that the IoT device performs a fake transmission, but the jammer does not attack the channel, the IoT device will then switch to the passive energy harvesting mode in the rest of time slot to save energy. In addition, it can be observed that the deception strategy is not always the best choice for the IoT device to perform as its efficiency depends largely on the IoT device's capacity, e.g., the current number of packets and energy levels. Thus, there are two critical factors which need to be defined in the state space for the IoT system, i.e., the energy levels in the battery and the number of packets waiting in the data buffer. Based on its current states, the IoT device can choose one possible action, observe its immediate result, and then continue improving its optimal policy through learning from its decisions. In reinforcement learning, $Q$-learning is the most effective method and widely adopted in the literature. The key idea of $Q$-learning method is using a $Q$-table to store all pairs of state-action values. Then, through the trial-and-error processes, the $Q$-values will be updated and it is proved in~\cite{Sutton1998Reinforcement} that if the agent has sufficient time to learn, the $Q$-values will finally converge to the optimal values. In other words, the IoT device will obtain the optimal policy, i.e., the policy which shows the best action to take give the current state, through this learning process.

It is also important to note that for reactive jamming attacks, the jammer only attacks the channel after the IoT makes an active transmission (either fake or actual transmission), and thus the channel status cannot be used as a decision factor in the state space (as the channel is always idle at the beginning of a time slot). As a result, given only local information, i.e.,the number of packets and energy levels, it may take very long time for the IoT device to obtain the optimal policy through trial-and-error processes. Thus, in the next section, we introduce a state-of-the-art technique, called DRL, which can obtain the optimal policy much faster than the conventional reinforcement learning method. This makes our proposed anti-jamming framework much more reliable and efficient to implement in practice.

\subsubsection{Deep Reinforcement Learning}

\begin{figure*}[!]
	\centering
	\includegraphics[width=\linewidth]{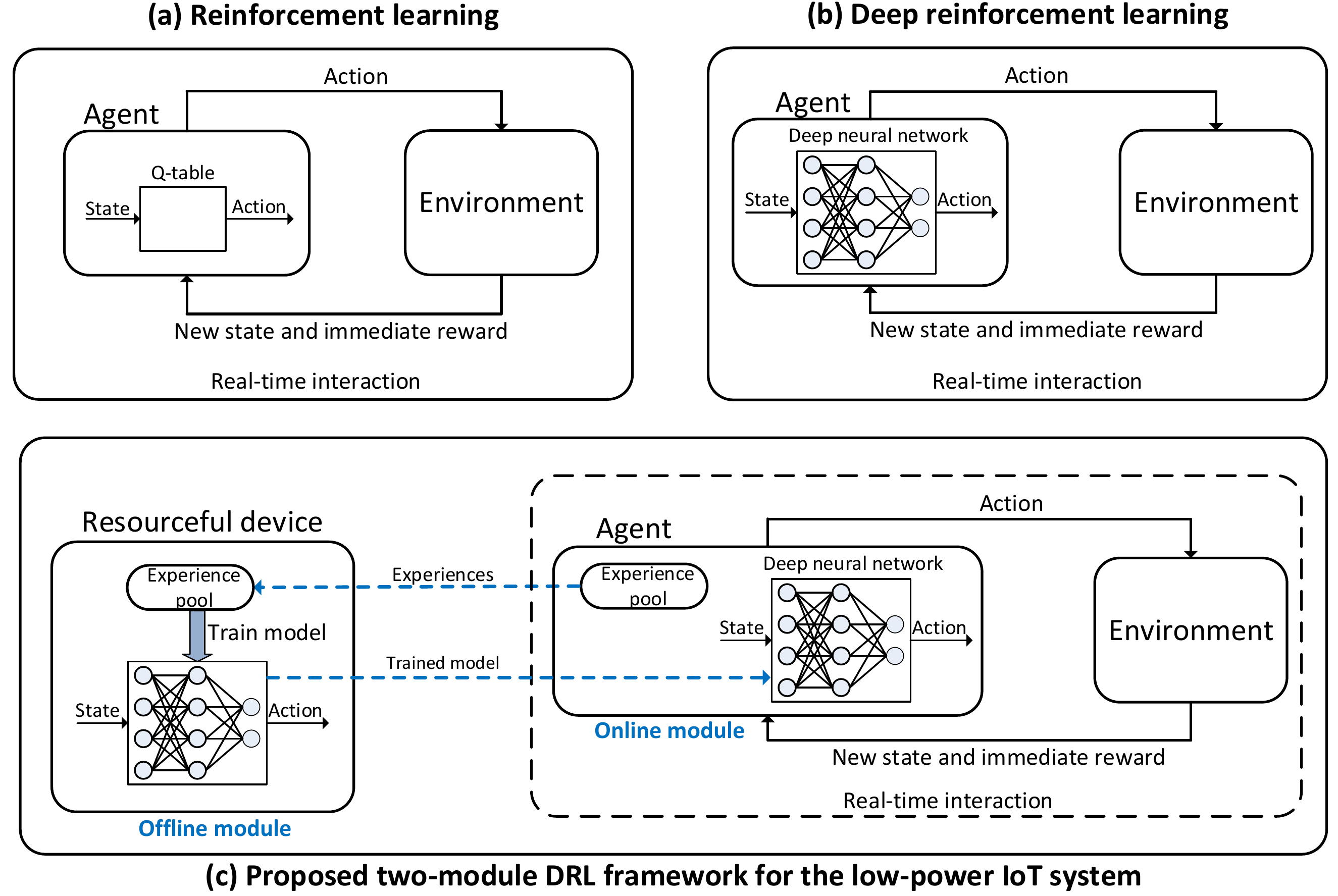}
	\caption{Illustrations of (a) Reinforcement learning, (b) Deep reinforcement learning, and (c) Proposed two-module DRL framework for the low-power IoT system.}
	\label{fig_RLvsDRL}
\end{figure*}

Recently, DRL has been emerging to be an enabling AI technology which can overcome limitations of reinforcement learning algorithms and making applications of reinforcement learning more and more efficiency in practice~\cite{Mnih2015Nature,Luong2019Applications}. The key idea of DRL is utilizing the advantage of deep artificial neural networks (DNN) to significantly improve the learning speed of trail-and-error processes. In particular, instead of making decisions based on the $Q$-values in the $Q$-table as in reinforcement learning, the agent in DRL uses a DNN to make decisions as illustrated in Fig.~\ref{fig_RLvsDRL}(b). A DNN is composed by an input layer, an output layer, and several hidden layers. Each layer has many neurons and the connection between two neurons is called weight. Then, if a state is sent to the DNN, features of the state will be extracted (e.g., the energy level and the number of packets waiting in the data buffer) and fed to the input layer. After that, based on values of weights, the DNN can find the optimal value, i.e., the best action for the IoT device to take, at the output layer.

Theoretically, reinforcement learning and DRL are similar because both of them use the same procedure, i.e., make a decision and then learn from results obtained from this decision. However, the key difference between reinforcement learning and DRL is at the way they train their models. Specifically, for a reinforcement learning process, each experience (defined to be a set of the current state, the action selected at this state, the reward received at this state after the action is executed, and the next state of the system) is learned only one time. This means that the decision policy of agent is updated immediately once an experience is generated and this experience might not be used in the future to update the decision policy. However, for the DRL, experiences will be stored in an experience pool and they will be combined randomly to train the DNN many times. In this case, the DNN can learn from a large set of different scenarios, and thus the learning quality is much better than that of the conventional reinforcement learning. 

Obviously, training processes are the key factors for DRL to significantly outperform conventional reinforcement learning algorithms, but they come with intensive computing costs. As a result, DRL algorithms might not be practically implemented on low-power devices, e.g., the IoT device considered in this paper. Therefore, we introduce a two-module DRL framework which can be efficiently implemented on the low-power IoT system. The first module, called online module, is implemented on the IoT device. This module includes two main components, i.e., an experience pool and a DNN as illustrated in Fig.~\ref{fig_RLvsDRL}(c). The DDN is used to help the IoT to choose the best action to take given the current state. The experiment pool is used to store all experiences which the IoT device obtains during the interaction process with the jammer. Experiences in the experience pool will be periodically, e.g., after every new 1000 experiences, sent to a resourceful device, e.g., the IoT gateway or a nearby computer, to train the DNN. After the DNN is trained, the newly updated model will be sent to the IoT device for updating the local model. In this way, all training processes are ``offloaded'' to the resourceful devices, and the IoT device only needs to run simple low-cost tasks during the real-time interaction with the jammer.

\subsection{Simulation Results}

In this section, we carry out simulations to evaluate the performance of our proposed DRL-based deception strategy framework to defeat reactive jamming attacks. We set the maximum data queue size and battery levels at 10 units. In particular, the battery can store up to 10 units of energy with each energy unit set to be 60$\mu J$~\cite{Papotto2014A90}, and the data buffer can store up to 10 packets with each packet size set to be 300 bits~\cite{Blasco2013Alearning}. At each time slot, the IoT device can choose to perform a fake transmission or an actual transmission which consumes one or three units of energy, respectively. If the IoT device chooses to perform an actual transmission and there is no attack on the channel, the IoT can successfully transmit three packets to the gateway. However, if the jammer attacks the channel, then all packets will be dropped. If the IoT device performs a fake transmission and the jammer attacks the channel, the IoT device can harvest three units of energy or backscatter one packet to the gateway. In addition, at each time slot, it is assumed that there are two packets generated at the IoT device with probability 0.5, and the IoT device can harvest one unit of energy from surrounding environment with probability 0.3. The jammer attack and packet arrival probabilities will be varied to evaluate the influence of the jammer and IoT device to the system, respectively. Parameters of DNN are adopted from general settings in the literature~\cite{Mnih2015Nature,Luong2019Applications}, e.g., two fully-connected hidden layers with 200 neurons per layer. In addition, to evaluate the efficiency of the proposed deception mechanism taking both energy harvesting and backscattering into considerations, we compare with three other strategies: (1) Deception then Backscatter (DB), (2) Deception then Harvest Energy (DH), and (3) Without Deception (WD). To make fair comparisons, for the first two strategies, their optimal policies are also obtained by the DRL. For the last strategy, i.e., WD, the IoT device will transmit data as long as it has data and sufficient energy.

\begin{figure*}[!]
	\begin{center}
		$\begin{array}{cc} 
		\epsfxsize=3.2 in \epsffile{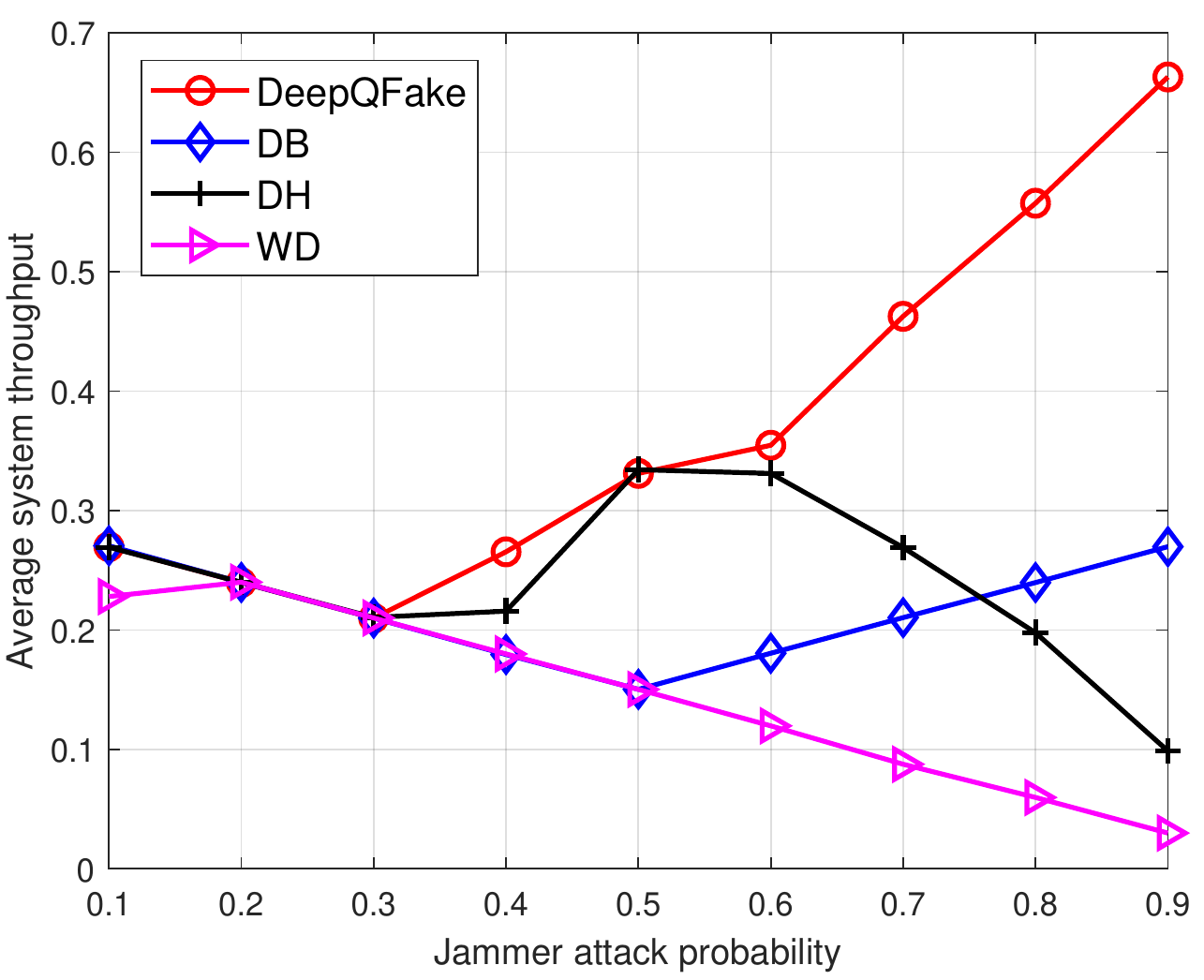} & 
		\epsfxsize=3.2 in \epsffile{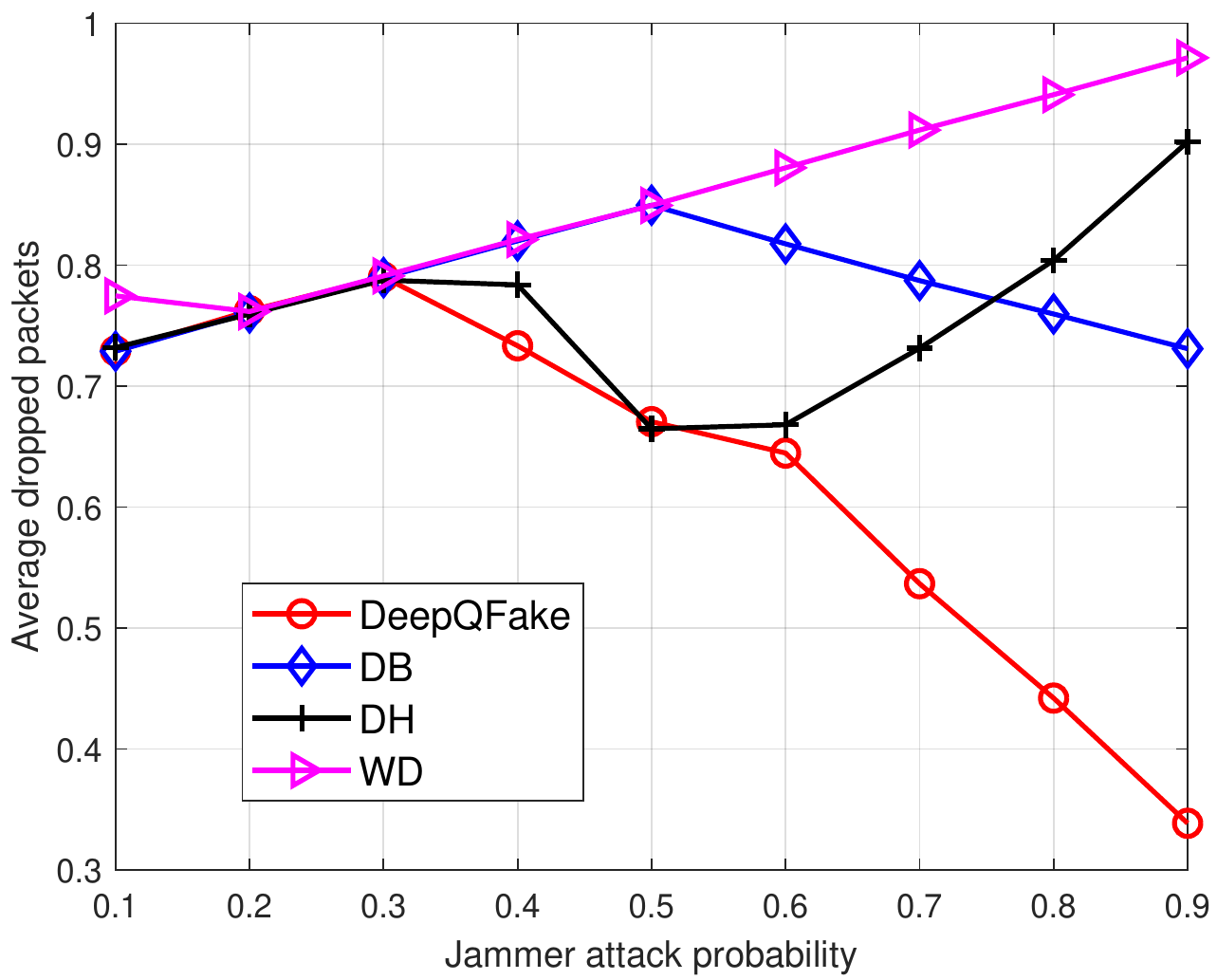} \\ [-0.2cm]
		(a) & (b) 
		\end{array}$
		\caption{Vary jamming attack probability.}
		\label{fig_Vary_jamming_attack_probability}
	\end{center}
\end{figure*}

First, we increase the probability of attacks (per time slot) from 0.1 to 0.9 and evaluate the performance of IoT system in terms of the average throughput and dropped packets under different anti-jamming strategies. As observed in Fig.~\ref{fig_Vary_jamming_attack_probability}(a), as the frequency of attacks increases from 0.1 to 0.3, the average throughput obtained by the proposed framework slightly drops from 0.28 to 0.21. However, interestingly, if we keep increasing the probability of the attack from 0.3 to 0.9, the average throughput gradually increases from 0.21 to 0.31 before soaring to 0.67 when the probability of attacks is at 0.9. The reason is that if the jammer often attacks the channel, it will unintentionally supply an abundant and strong RF signal resource for the IoT device. As a result, the IoT device can utilize such signals to enhance the performance for IoT system. Compared with other deception strategies, our proposed solution can simultaneously optimize backscattering, energy harvesting, and active transmissions, and thus it always achieves the best performance. In addition, it can be observed that without using the deception strategy, the throughput of IoT system will be dropped quickly as the probability of attacks increases.

\begin{figure*}[!]
	\begin{center}
		$\begin{array}{cc} 
		\epsfxsize=3.2 in \epsffile{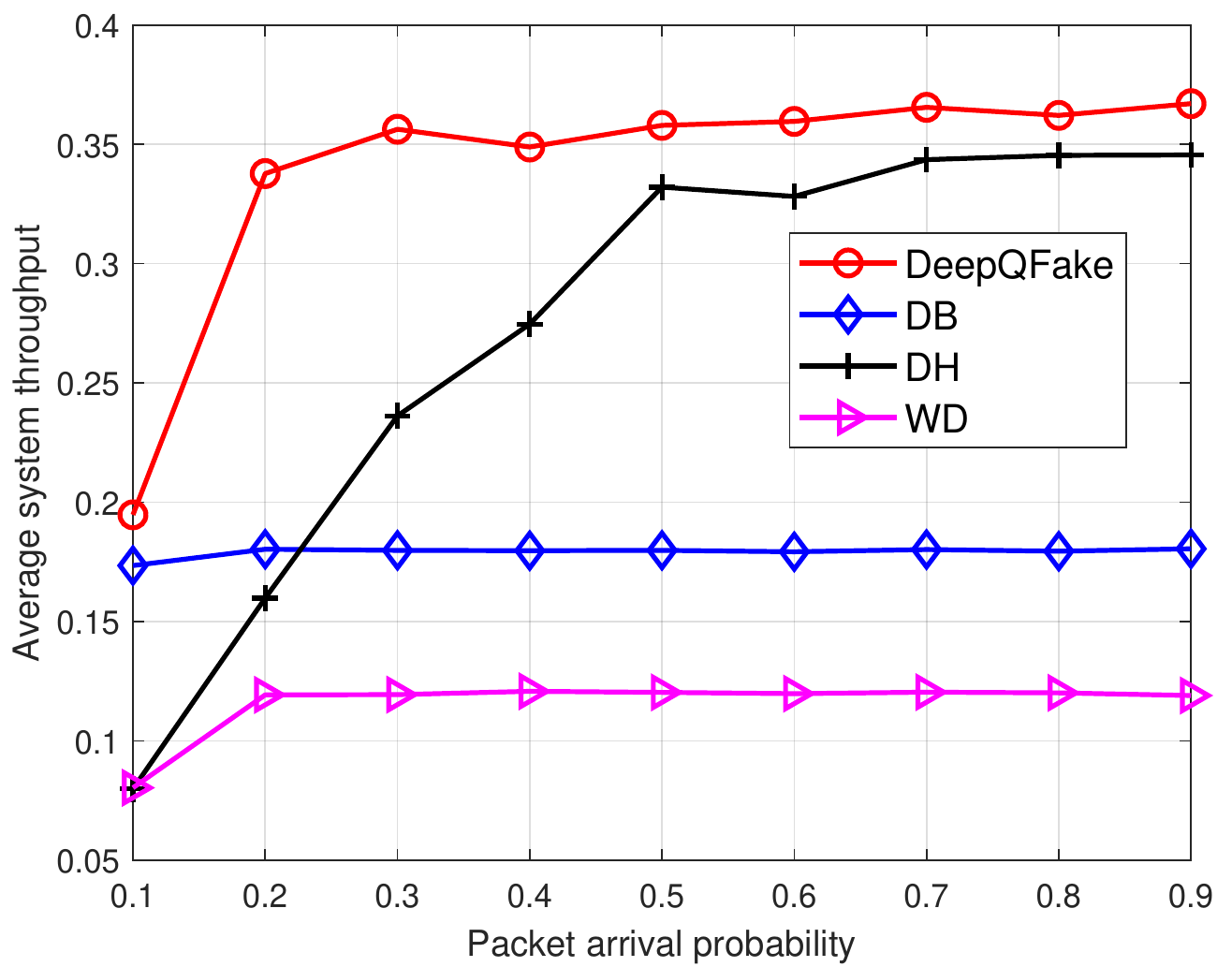} & 
		\epsfxsize=3.2 in \epsffile{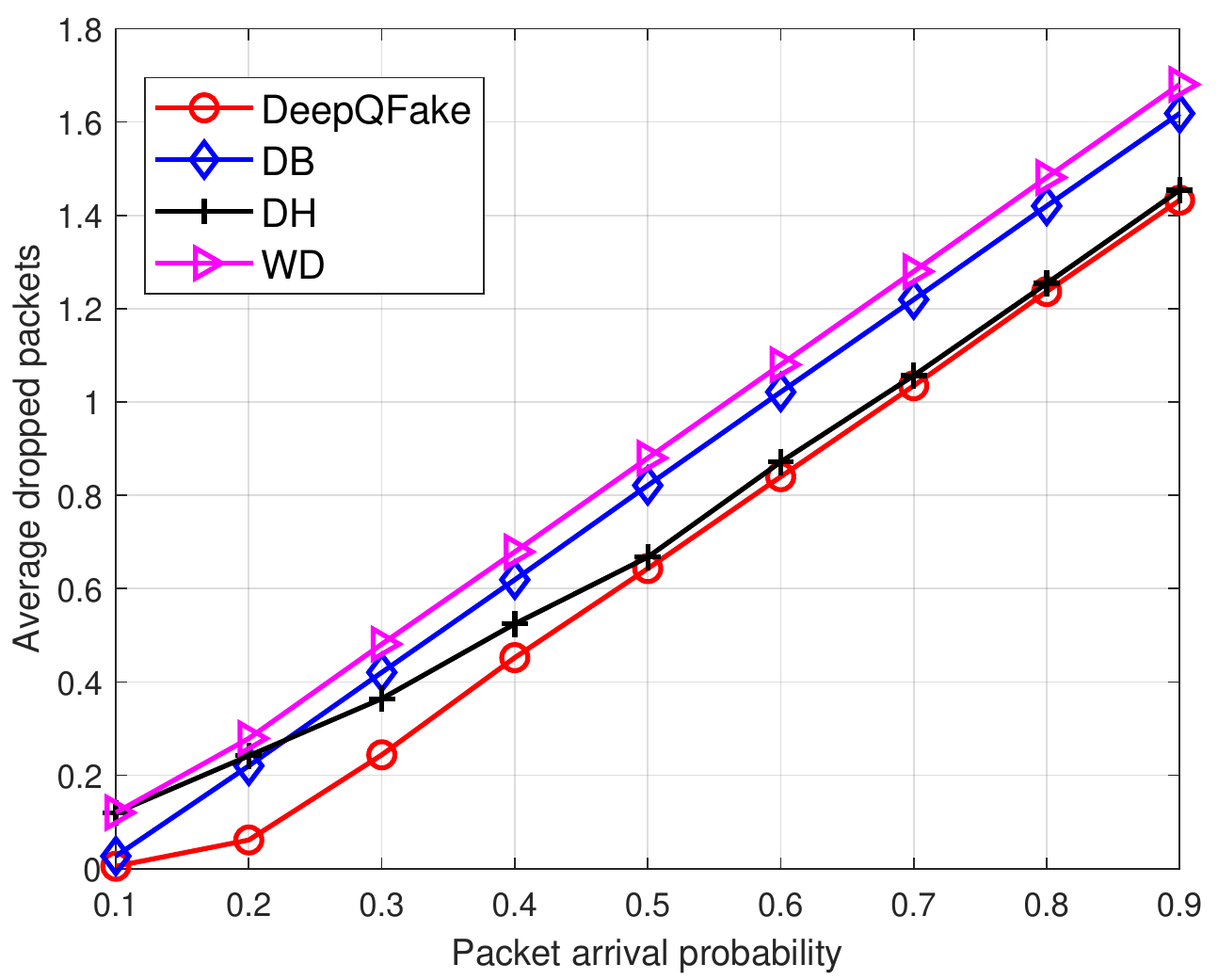} \\ [-0.2cm]
		(a) & (b) 
		\end{array}$
		\caption{Vary packet arrival probability.}
		\label{fig_Vary_packet_arrival_probability}
	\end{center}
\end{figure*}

In Fig.~\ref{fig_Vary_packet_arrival_probability}, we fix the probability of jamming attacks at 0.6 and vary the packet arrival probability to evaluate the IoT system performance. As the packet arrival probability increases from 0.1 to 0.2, the system throughput obtained by the proposed solution surges nearly 75\%, from 0.2 to 0.35. Nevertheless, if we keep increasing the packet arrival probability, the system throughput only slightly increases because it reaches the saturation state, i.e., the IoT system cannot improve its performance given the current system setting. At the saturation point, the average throughput obtained by the proposed solution can achieve 6\%, 94\%, and 170\% greater than that of the DH, DB, and WD, respectively.

\section{Conclusion}
\label{sec:Sum}

In this paper, we have introduced the novel anti-jamming framework for the low-power IoT system. This framework is developed based on the intelligent deception strategy and the low-cost DRL framework. The proposed deception strategy is a groundbreaking solution in dealing with reactive jamming attacks, meanwhile the proposed two-module DRL is an AI-based effective learning tool to quickly find an optimal defense policy for the IoT system given the incomplete information from the jammer and the dynamic of the wireless environment. Through simulations, we have clearly shown the efficiency as well as outperformance of proposed framework compared with other anti-jamming strategies.



\begin{thebibliography}{100}
\bibliographystyle{IEEEtranS}

\bibitem{Fuqaha2015Internet}
A.~A.~Fuqaha, M.~Guizani, M.~Mohammadi, M.~Aledhari, and M.~Ayyash, ``Internet of things: A survey on enabling technologies, protocols, and applications,'' \emph{IEEE Communications Surveys \& Tutorials}, vol. 17, no. 4, pp. 2347-2376, Jun. 2015.

%

\bibitem{Manchester}
Manchester car park lock hack leads to horn-blare hoo-ha. Available Online: \url{https://www.theregister.co.uk/2015/05/20/car_park_vehicle_locks_hacked_en_masse}. Last Accessed on Oct. 2019.

\bibitem{Huynh2018Ambient}
N.~V.~Huynh, D.~T.~Hoang, X.~Lu, D.~Niyato, P.~Wang, and D.~I.~Kim, ``Ambient backscatter communications: A contemporary survey,'' \emph{IEEE Communications Surveys \& Tutorials}, vol. 20, no. 4, pp. 2889-2922, Jan.
2018.

\bibitem{Yuan2013Coexistence}
W.~Yuan, X.~Wang, J.~P.~Linnartz, and I.~G.~Niemegeers, ``Coexistence performance of IEEE 802.15. 4 wireless sensor networks under IEEE 802.11 b/g interference,'' \emph{Wireless Personal Communications}, vol. 68, no. 2, pp. 281-302, Jan. 2013.

\bibitem{Mpitziopoulos2009Asurvey}
A.~Mpitziopoulos, D.~Gavalas, C.~Konstantopoulos, and G.~Pantziou, ``A survey on jamming attacks and countermeasures in WSN,'' \emph{IEEE Communications Surveys \& Tutorials}, vol. 11, no. 4, pp. 42-56, Dec. 2009.

\bibitem{HuynhJam2019}
N.~V.~Huynh, D.~N.~Nguyen, D.~T.~Hoang, E.~Dutkiewicz, ``Jam me if you can: Defeating jammer with deep dueling neural network architecture and ambient backscattering augmented communications,'' \emph{IEEE Journal on Selected Areas in Communications}, Early Access, Aug. 2019.

\bibitem{Liu2013Ambient}
V. Liu, A. Parks, V. Talla, S. Gollakota, D. Wetherall, and J. R. Smith, ``Ambient backscatter: Wireless communication out of thin air,'' in \emph{ACM SIGCOMM}, Hong Kong, Aug. 2013, pp. 39-50.

\bibitem{Kimionis2014Increased}
J.~Kimionis, A.~Bletsas, and J.~N.~Sahalos, ``Increased range bistatic scatter radio,'' \emph{IEEE Transactions on Communications}, vol. 62, no.3, pp. 1091-1104, Feb. 2014.

\bibitem{Sciancalepore2018Strength}
S.~Sciancalepore, G.~Oligeri, and R.~D.~Pietro, ``Strength of crowd (SOC) - Defeating a reactive jammer in IoT with decoy messages,'' \emph{Sensors}, vol. 18, no. 10, pp. 3492, Oct. 2018.

\bibitem{Xu2012Joint}
K.~Xu, Q.~Wang, and K.~Ren, `` Joint UFH and power control for effective wireless anti-jamming communication,'' in \emph{INFOCOM}, Orlando, FL, USA, Mar. 2012, pp. 738-746.

\bibitem{Luong2019Applications}
N.~C.~Luong, D.~T.~Hoang, S.~Gong, D.~Niyato, P.~Wang, Y.-C.~Liang, and D.~I.~Kim, ``Applications of deep reinforcement learning in communications and networking: A survey,'' \emph{IEEE Communications Surveys and Tutorials}, Early Access, May 2019.

\bibitem{Sutton1998Reinforcement}
R.~S.~Sutton and A.~G.~Barto, \emph{Reinforcement learning: An introduction}.
MIT press Cambridge, 1998.

\bibitem{Mnih2015Nature}
V. Mnih, et al., ``Human-level control through deep reinforcement learning,'' \emph{Nature}, vol. 518, no. 7540, pp. 529-533, Feb. 2015.

\bibitem{Papotto2014A90}
G. Papotto et al., ``A 90-nmCMOS 5-Mbps crystal-Less RF-powered transceiver for wireless sensor network nodes,'' \emph{IEEE Journal of Solid-State Circuits}, vol. 49, no. 2, pp. 335-346, Feb. 2014.

\bibitem{Blasco2013Alearning}
P.~Blasco, D.~Gunduz, and M.~Dohler, ``A learning theoretic approach to energy harvesting communication system optimization,'' \emph{IEEE Transactions	Wireless Communications}, vol. 12, no. 4, pp. 1872-1882, Apr. 2013.

\end{thebibliography}
\end{document}